\documentclass{PoS}
\usepackage{psfrag,epsfig,graphicx,graphics,xcolor}
\usepackage{wrapfig}
\usepackage{amsmath}

\def\slashchar#1{\setbox0=\hbox{$#1$}
   \dimen0=\wd0
   \setbox1=\hbox{/} \dimen1=\wd1
   \ifdim\dimen0>\dimen1
      \rlap{\hbox to \dimen0{\hfil/\hfil}}
      #1
   \else
      \rlap{\hbox to \dimen1{\hfil$#1$\hfil}}
      /
   \fi}

\def\bei{\begin{itemize}}
\def\ei{\end{itemize}}

\def\beeq{\begin{eqnarray}} 
\def\beqa{\begin{eqnarray}}
\def\bea{\begin{eqnarray}}

\def\eea{\end{eqnarray}}
\def\eqa{\end{eqnarray}}
\def\eeeq{\end{eqnarray}}

\def\eqar{\end{array}}
\def\beqar{\begin{array}}

\def\beas{\begin{eqnarray*}}
\def\beqas{\begin{eqnarray*}}

\def\eqas{\end{eqnarray*}}
\def\eeas{\end{eqnarray*}}

\def\beq{\begin{equation}} 
\def\be{\begin{equation}}

\def\ee{\end{equation}}
\def\eq{\end{equation}}
\def\eeq{\end{equation}}

\def\beqd{\begin{displaymath}}
\def\eeqd{\end{displaymath}}
\def\eqd{\end{displaymath}}

\def\beeq{\begin{eqnarray}} \def\eeeq{\end{eqnarray}}

\newcommand{\fin}{\end{document}}

\newcommand{\veck}{{\bf k}}

\newcommand{\veckone}{{\bf k}_1}
\newcommand{\vecktwo}{{\bf k}_2}
\newcommand{\veckj}{{\bf k}_{J}}

\newcommand{\veckjone}{{\bf k}_{J,1}}
\newcommand{\veckjtwo}{{\bf k}_{J,2}}

\newcommand{\deins}[1]{{\rm d}#1\,}
\newcommand{\dzwei}[1]{{\rm d}^2#1\,}
\newcommand{\dk}{\dzwei{\veck}}

\newcommand{\dkone}{\dzwei{\veckone}}
\newcommand{\dktwo}{\dzwei{\vecktwo}}

\newcommand{\dsigma}{\deins{\sigma}}
\newcommand{\dsigmahat}{\deins{{\hat\sigma}_{\rm{ab}}}}
\newcommand{\dnu}{\deins{\nu}}

\newcommand{\dx}{\deins{x}}
\newcommand{\dxone}{\deins{x_1}}
\newcommand{\dxtwo}{\deins{x_2}}
\newcommand{\dyjetone}{\deins{y_{J,1}}}
\newcommand{\dyjettwo}{\deins{y_{J,2}}}
\newcommand{\dphij}{\deins{\phi_{J}}}
\newcommand{\dphijone}{\deins{\phi_{J,1}}}
\newcommand{\dphijtwo}{\deins{\phi_{J,2}}}
\newcommand{\dtwojets}{{\rm d}|\veckjone|\,{\rm d}|\veckjtwo|\,\dyjetone \dyjettwo}

\newcommand{\shat}{{\hat s}}

\newcommand{\asbar}{{\bar{\alpha}}_s}

\newcommand{\avgcosn}{\langle \cos n \varphi \rangle}
\newcommand{\avgcos}{\langle \cos \varphi \rangle}
\newcommand{\avgcostwo}{\langle \cos 2 \varphi \rangle}

\title{Mueller Navelet jets at LHC: a clean test of QCD resummation effects at high energy?}

\ShortTitle{Mueller Navelet jets at LHC: a clean test of QCD resummation effects at high energy?}

\author{\speaker{Bertrand Duclou\'e}\\
        LPT, Universit\'e Paris-Sud, CNRS, 91405, Orsay, France\\
        E-mail: \email{Bertrand.Ducloue@th.u-psud.fr}}

\author{Lech Szymanowski\\
        National Center for Nuclear Research (NCBJ), Warsaw, Poland\\
        E-mail: \email{Lech.Szymanowski@fuw.edu.pl}}

\author{Samuel Wallon\\
        LPT, Universit{\'e} Paris-Sud, CNRS, 91405, Orsay, France\\
        UPMC Univ. Paris 06, facult\'e de physique, 4 place Jussieu, 75252 Paris Cedex 05, France
        E-mail: \email{Samuel.Wallon@th.u-psud.fr}}

\abstract{Mueller Navelet jets were proposed more than 25 years ago as a decisive test of BFKL dynamics at hadron colliders. We here present a complete next-to-leading BFKL study of the azimuthal decorrelation of these jets. This includes both next-to-leading corrections to the Green's function and next-to-leading corrections to the jet vertices. We compare our results with recent data taken at the LHC and results obtained in a fixed order next-to-leading-order (NLO) calculation.}

\FullConference{XXI International Workshop on Deep-Inelastic Scattering and Related Subjects - DIS2013\\
		 22-26 April, 2013\\
		 Marseille, France}

\begin{document}

\section{Introduction}
Several observables have been suggested as a way to study the high energy limit of QCD. In this limit, the smallness of the strong coupling $\alpha_s$ can be compensated by large logarithmic enhancements of the type $[\alpha_s\ln(s/|t|)]^n$ which have to be resummed, giving rise to the 
leading logarithmic (LL) Balitsky-Fadin-Kuraev-Lipatov (BFKL)  Pomeron \cite{Fadin:1975cb,Kuraev:1976ge,Kuraev:1977fs,Balitsky:1978ic}. Mueller and Navelet proposed to study the production of two jets with large rapidity separation at hadron colliders \cite{Mueller:1986ey}. In a pure leading order collinear treatment these two jets would be emitted back to back, while a BFKL treatment allows some emission between these jets and so should lead to a larger cross section and lower angular correlation of the jets. We present results of a full NLL analysis where the NLL corrections are included for the BFKL Green's function~\cite{Fadin:1998py,Ciafaloni:1998gs} and the jet vertices~\cite{Bartels:2001ge,Bartels:2002yj,Caporale:2011cc}.

Here we will focus on the azimuthal correlations $\avgcosn$ and ratios of these observables at a center of mass energy $\sqrt{s}=7$ TeV which have been measured recently at the LHC by the CMS collaboration~\cite{CMS-PAS-FSQ-12-002} and make some comparison of our results~\cite{Ducloue:2013hia} both with these data and with results obtained in a fixed order NLO treatment.

\section{Basic formulas}

\begin{figure}[htbp]
\centering
\includegraphics[height=8cm]{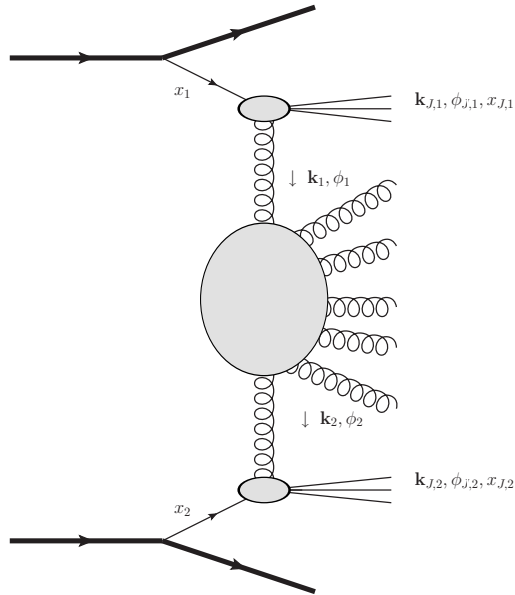}
\caption{Kinematics of the process}
\label{Fig:kinematics}
\end{figure}

We consider the process shown on figure~\ref{Fig:kinematics}, in which two hadrons collide at a center of mass energy $\sqrt{s}$. Using collinear factorization, the differential cross section reads
\begin{equation}
  \frac{\dsigma}{\dtwojets} = \sum_{{\rm a},{\rm b}} \int_0^1 \dxone \int_0^1 \dxtwo f_{\rm a}(x_1) f_{\rm b}(x_2) \frac{\dsigmahat}{\dtwojets},
\end{equation}
where $\veckjone$, $\veckjtwo$ are the transverse momenta of the jets, $y_{J,1}$ and $y_{J,2}$ their rapidities and $f_{\rm a,\, b}$ are the parton distribution functions (PDFs) of a parton a (b) in the according proton. In this expression, the partonic cross section is
\begin{equation}
  \frac{\dsigmahat}{\dtwojets} = \int \dphijone\dphijtwo\int\dkone\dktwo V_{\rm a}(-\veckone,x_1)\,G(\veckone,\vecktwo,\shat)\,V_{\rm b}(\vecktwo,x_2),\label{eq:bfklpartonic}
\end{equation}
where $\phi_{J,1}$ and $\phi_{J,2}$ are the azimuthal angles of the jets, $V_{a,b}$ is the jet vertex initiated by the parton a (b) and $G$ is the BFKL Green's function which depends on $\shat=x_1 x_2 s$.
It is convenient to introduce the coefficients $\mathcal{C}_n$, defined as
\begin{equation}
  \mathcal{C}_n = (4-3\delta_{n,0}) \int \dnu C_{n,\nu}(|\veckjone|,x_{J,1})C^*_{n,\nu}(|\veckjtwo|,x_{J,2}) \left( \frac{\shat}{s_0} \right)^{\omega(n,\nu)}\,,
  \label{Cn}
\end{equation}
such that
\begin{equation}
  \frac{\dsigma}{\dtwojets} = \mathcal{C}_0 \quad {\rm and} \quad 
  \langle\cos(n\varphi)\rangle \equiv \langle\cos\big(n(\phi_{J,1}-\phi_{J,2}-\pi)\big)\rangle = \frac{\mathcal{C}_n}{\mathcal{C}_0} .
\end{equation}
In eq.~(\ref{Cn}), $C_{n,\nu}$ is defined as
\begin{equation}
   C_{n,\nu}(|\veckj|,x_{J})= \int\dphij\dk \dx f(x) V(\veck,x) E_{n,\nu}(\veck) \cos(n\phi_J)\,,
  \label{Cnnu}
\end{equation}
with the LL BFKL eigenfunctions $E_{n,\nu}$ being
\begin{equation}
  E_{n,\nu}(\veckone) = \frac{1}{\pi\sqrt{2}}\left(\veckone^2\right)^{i\nu-\frac{1}{2}}e^{in\phi_1}\,.
\label{def:eigenfunction}
\end{equation}
At LL accuracy, $\omega(n,\nu)$ is
\begin{equation}
  \omega(n,\nu) = \asbar \chi_0\left(|n|,\frac{1}{2}+i\nu\right), \quad \chi_0(n,\gamma) = 2\Psi(1)-\Psi\left(\gamma+\frac{n}{2}\right)-\Psi\left(1-\gamma+\frac{n}{2}\right)\,,
\end{equation}
with $\asbar = \alpha N_c/\pi$ and $\Psi(z) = \Gamma'(z)/\Gamma(z)\,,$
and the vertex is
\begin{equation}
  V_{\rm a}(\veck,x)=V_{\rm a}^{(0)}(\veck,x) = \frac{\alpha_s}{\sqrt{2}}\frac{C_{A/F}}{\veck^2} \delta\left(1-\frac{x_J}{x}\right)|\veckj|\delta^{(2)}(\veck-\veckj)\,,
\end{equation}
($C_A$ for ${\rm a}={\rm g}$ and $C_F$ for ${\rm a}={\rm q}$), while at NLL, we have
\begin{equation}
  \omega(n,\nu) = \asbar \chi_0\left(|n|,\frac{1}{2}+i\nu\right) + \asbar^2 \left[ \chi_1\left(|n|,\frac{1}{2}+i\nu\right)-\frac{\pi b_0}{N_c}\chi_0\left(|n|,\frac{1}{2}+i\nu\right) \ln\frac{|\veckjone|\cdot|\veckjtwo|}{\mu_R^2} \right]\,,
\end{equation}
with $b_0=(33 - 2 \,N_f)/(12 \pi)$
and $V_{\rm a}(\veck,x) = V^{(0)}_{\rm a}(\veck,x) + \alpha_s V^{(1)}_{\rm a}(\veck,x)$. The expression of the NLL corrections to the Green's function resulting in $\chi_1$ can be found in eq.~(2.17) of ref.~\cite{Ducloue:2013hia}. The expressions of the NLL corrections to the jet vertices are quite lenghty and will not be reproduced here. They can be found in ref.~\cite{Colferai:2010wu}, as extracted from refs.~\cite{Bartels:2001ge,Bartels:2002yj} after correcting a few misprints of ref.~\cite{Bartels:2001ge}.  They have been recently reobtained in ref.~\cite{Caporale:2011cc}. In the limit of small cone jets, they have been computed in ref.~\cite{Ivanov:2012ms} and applied to phenomenology in refs.~\cite{Caporale:2012ih,Caporale:2013uva}.
Here we will use the cone algorithm with a size of $R_{\rm cone}=0.5$. We choose the central value $\sqrt{|\veckjone|\cdot |\veckjtwo|}$ for the renormalization scale $\mu_R$, the factorization scale $\mu_F$ and the energy scale $\sqrt{s_0}$, and vary these scales by a factor of 2 to estimate the scale uncertainty of our calculation. We use the MSTW 2008 PDFs \cite{Martin:2009iq} and a two-loop running coupling. We also include collinear improvement to the Green's function as was suggested in refs.~\cite{Salam:1998tj,Ciafaloni:1998iv,Ciafaloni:1999yw,Ciafaloni:2003rd} and extended for $n \neq 0$ in refs.~\cite{Vera:2007kn,Schwennsen:2007hs,Marquet:2007xx}.

\section{Results: symmetric configuration}

In this section, we show results for a symmetric configuration (identical lower cut for the transverse momenta of the jets) with cuts
\begin{eqnarray}
 35\,{\rm GeV} < &|\veckjone|,|\veckjtwo|& < 60 \,{\rm GeV} \,, \nonumber\\
 0 < &y_1, \, y_2& < 4.7\,.
 \label{sym-cuts}
\end{eqnarray}
This is close to the cuts used by CMS in~\cite{CMS-PAS-FSQ-12-002} with the exception that for numerical reasons we have to set an upper cut on the transverse momenta of the jets. We have checked that our results do not depend strongly on the value of this cut as the cross section is strongly peaked near the minimum value allowed for $\veckjone$ and $\veckjtwo$. This enables us to compare our predictions with LHC data.

In the following we will study several BFKL scenarios, from a pure LL approximation (LL Green's function and leading order jet vertex) to a full NLL calculation (NLL Green's function and NLL jet vertex). The color convention we will use for the plots showing the different approaches is the following:
\begin{equation}
\hspace{-.3cm}
\begin{tabular}{ll}
blue: &  pure LL result \\
magenta: &   combination of LL vertices with  pure NLL Green's function \\
green: &  combination of LL vertices with collinear improved NLL Green's function \\
brown: &  pure NLL result \\
red: &  full NLL vertices with  collinear improved NLL Green's function.
\end{tabular}
\label{def:colors}
\end{equation}

\begin{figure}[htbp]
  \def\sca{.6}
  \psfrag{central}[l][r][0.5]{\hspace{-0.9cm}pure NLL}
  \psfrag{muchange_0.5}[l][r][\sca]{\hspace{-1.8cm}\footnotesize $\mu_F \to \mu_F/2$}
  \psfrag{muchange_2.0}[l][r][\sca]{\hspace{-1.9cm} \footnotesize $\mu_F \to2 \mu_F$}
  \psfrag{s0change_0.5}[l][r][\sca]{\hspace{-1.95cm} \footnotesize $\sqrt{s_0} \to \sqrt{s_0}/2$}
  \psfrag{s0change_2.0}[l][r][\sca]{\hspace{-1.95cm} \footnotesize $\sqrt{s_0} \to 2 \sqrt{s_0}$}
  \psfrag{CMS}[l][r][0.5]{\hspace{-0.7cm} CMS data}
  \psfrag{cos}{\raisebox{.1cm}{\scalebox{0.9}{$\langle \cos \varphi\rangle$}}}
  \psfrag{Y}{\scalebox{0.9}{$Y$}}
  \begin{minipage}{0.49\textwidth}
      \includegraphics[width=7cm]{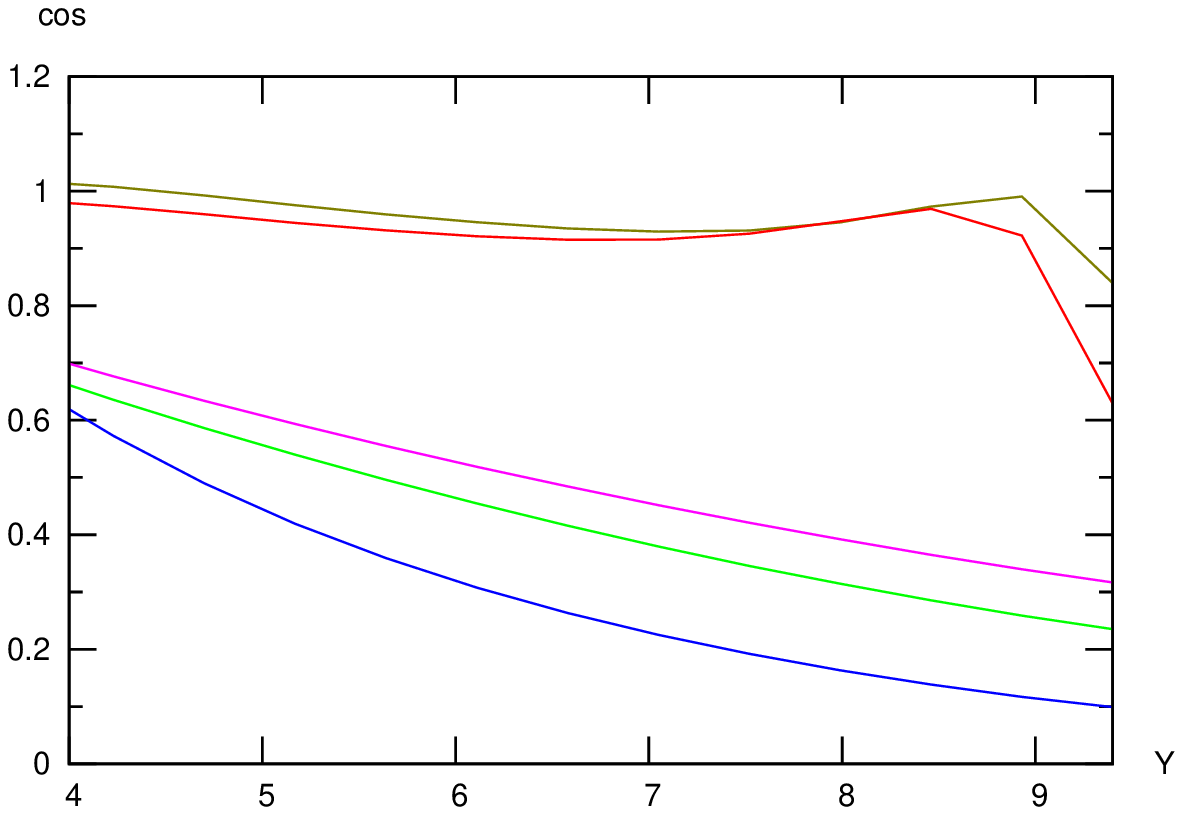}
  \end{minipage}
  \begin{minipage}{0.49\textwidth}
      \includegraphics[width=7cm]{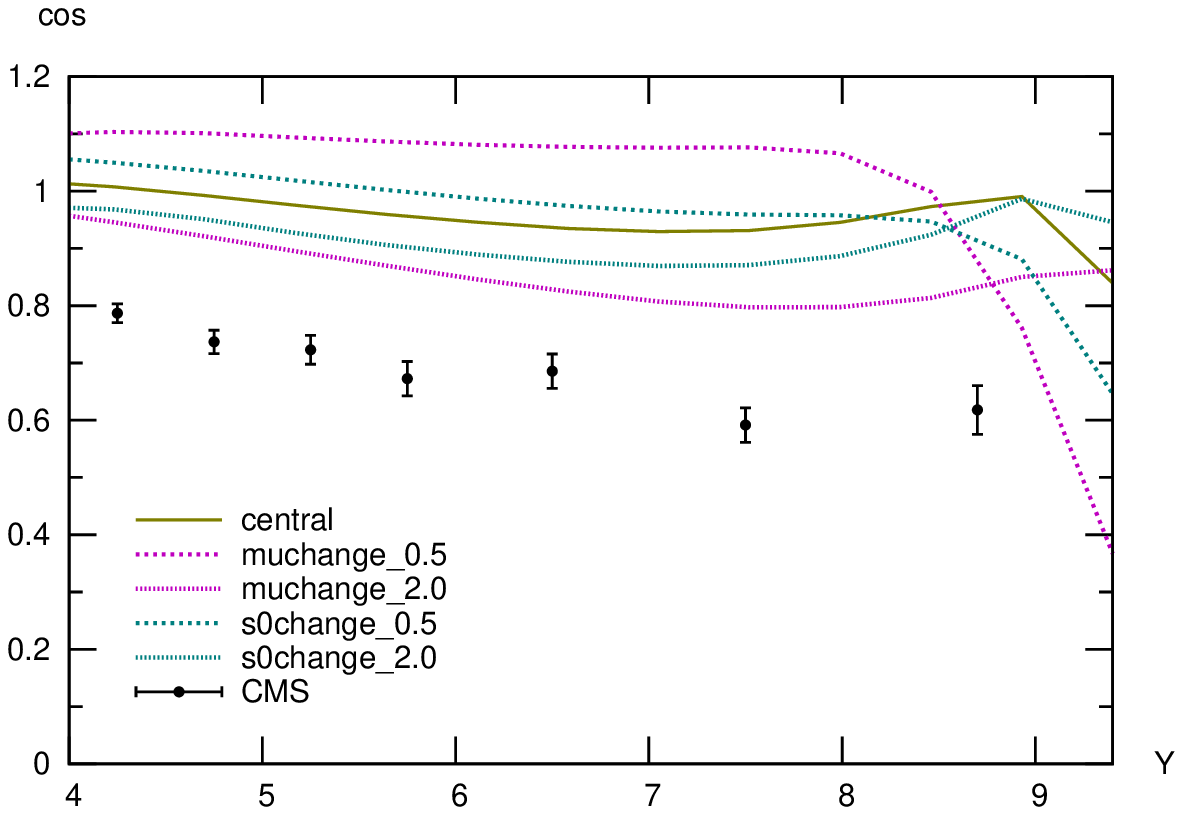}
  \end{minipage}
  \caption{Left: value of $\avgcos$ as a function of the rapidity separation $Y$, using symmetric cuts defined in (\protect\ref{sym-cuts}), for the 5 different BFKL treatments (\protect\ref{def:colors}). Right: comparison of the full NLL BFKL calculation including the scale uncertainty with CMS data.}
  \label{Fig:cos_sym}
\end{figure}

We begin our analysis with the azimuthal correlation $\avgcos$. In figure~\ref{Fig:cos_sym}~(L) we show the variation of $\avgcos$ with respect to the rapidity separation between the two jets $Y$ in the 5 scenarios (\ref{def:colors}). We observe that a pure LL treatment leads to a large decorrelation between the two jets, and that the NLL corrections to the Green's function restores a little correlation. The effect of the NLL corrections to the jet vertices is even larger and so a full NLL treatment predicts a value of $\avgcos$ very close to 1 (i.e. very close to be back-to-back). We also note that in this case the collinear improvement of the Green's function has a very small effect compared to the LL vertices case.
In figure~\ref{Fig:cos_sym}~(R) we show the variation of our NLL result when varying $\mu$ and $s_0$ by a factor of 2 and compare it with CMS data (black dots with error bars). We see that NLL BFKL predicts a larger correlation than seen in the data, but this observable is strongly dependent on the value of the scales.

\begin{figure}[htbp]
  \def\sca{.6}
  \psfrag{central}[l][r][0.5]{\hspace{-2cm}pure NLL}
  \psfrag{muchange_0.5}[l][r][\sca]{\hspace{-1.7cm}\footnotesize $\mu_F \to \mu_F/2$}
  \psfrag{muchange_2.0}[l][r][\sca]{\hspace{-1.8cm} \footnotesize $\mu_F \to2 \mu_F$}
  \psfrag{s0change_0.5}[l][r][\sca]{\hspace{-1.95cm} \footnotesize $\sqrt{s_0} \to \sqrt{s_0}/2$}
  \psfrag{s0change_2.0}[l][r][\sca]{\hspace{-1.95cm} \footnotesize $\sqrt{s_0} \to 2 \sqrt{s_0}$}
  \psfrag{CMS}[l][r][0.5]{\hspace{-2.1cm} CMS data}
  \psfrag{cos}{\raisebox{.1cm}{\scalebox{0.9}{$\langle \cos 2 \varphi\rangle$}}}
  \psfrag{Y}{\scalebox{0.9}{$Y$}}
  \begin{minipage}{0.49\textwidth}
      \includegraphics[width=7cm]{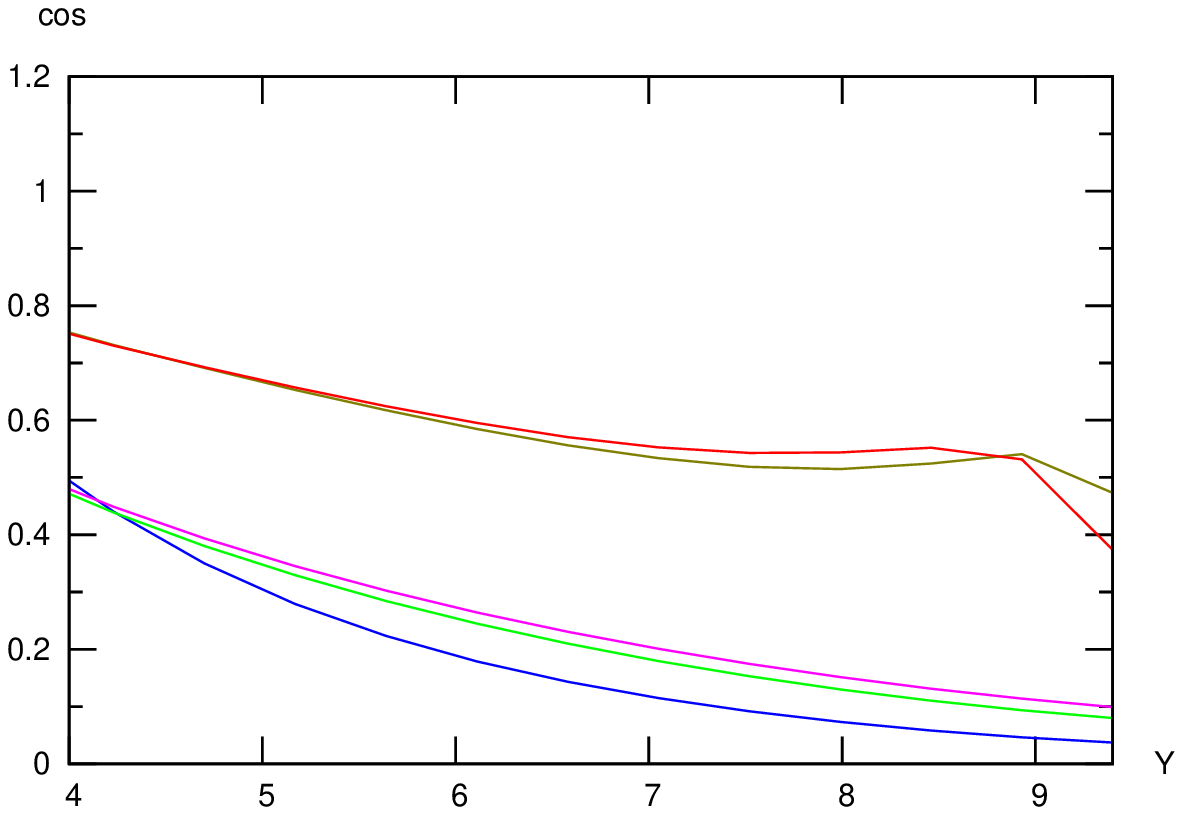}
  \end{minipage}
  \begin{minipage}{0.49\textwidth}
      \includegraphics[width=7cm]{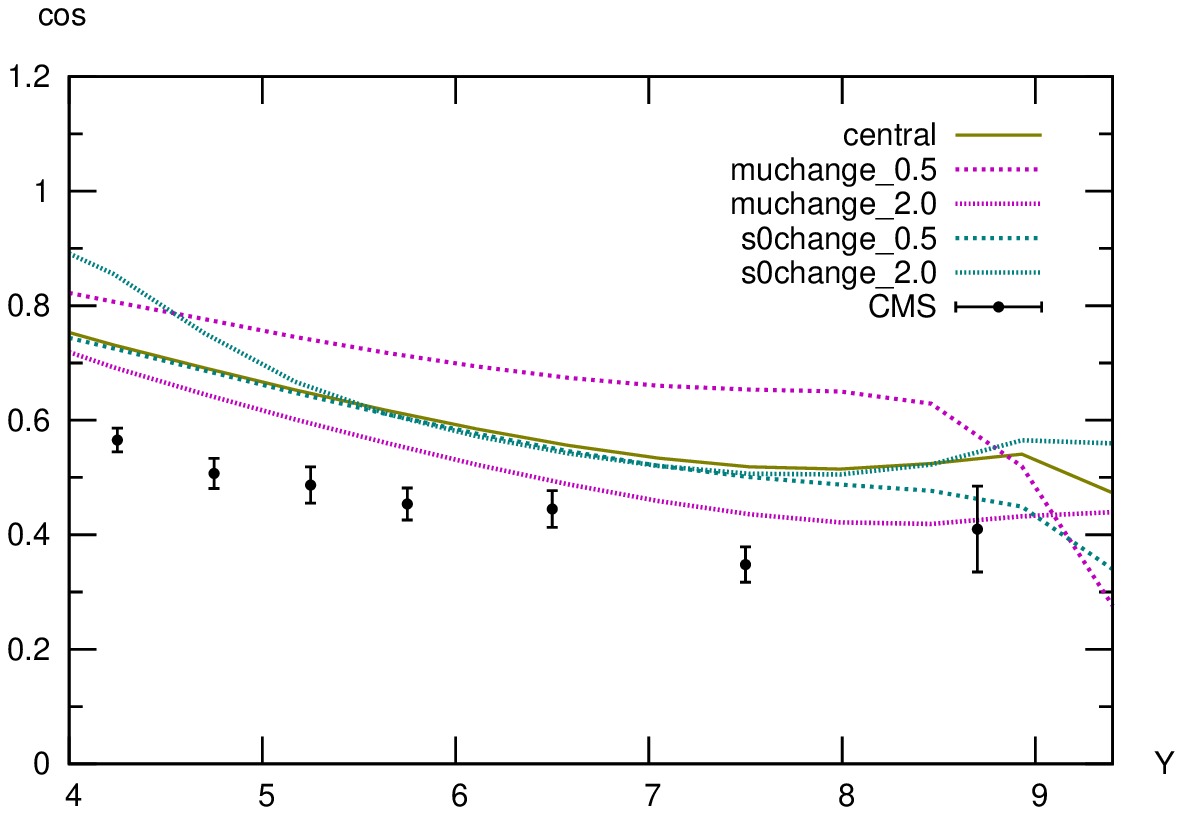}
  \end{minipage}
  \caption{Left: value of $\avgcostwo$ as a function of the rapidity separation $Y$, using symmetric cuts defined in (\protect\ref{sym-cuts}), for the 5 different BFKL treatments (\protect\ref{def:colors}). Right: comparison of the full NLL BFKL calculation including the scale uncertainty with CMS data.}
  \label{Fig:cos2_sym}
\end{figure}

In figure~\ref{Fig:cos2_sym} we consider the observable $\avgcostwo$. We observe a similar behavior as for $\avgcos$, the NLL corrections to the jet vertices produce a much larger change than the NLL corrections to the Green's function. Again the dependence of the NLL calculation on the choice of $\mu$ and $s_0$ is large, and taking this dependence into account the BFKL result is not very far from data.

\begin{figure}[htbp]
  \def\sca{.6}
  \psfrag{central}[l][r][0.5]{\hspace{-0.9cm}pure NLL}
  \psfrag{muchange_0.5}[l][r][\sca]{\hspace{-1.8cm}\footnotesize $\mu_F \to \mu_F/2$}
  \psfrag{muchange_2.0}[l][r][\sca]{\hspace{-1.9cm} \footnotesize $\mu_F \to2 \mu_F$}
  \psfrag{s0change_0.5}[l][r][\sca]{\hspace{-1.95cm} \footnotesize $\sqrt{s_0} \to \sqrt{s_0}/2$}
  \psfrag{s0change_2.0}[l][r][\sca]{\hspace{-1.95cm} \footnotesize $\sqrt{s_0} \to 2 \sqrt{s_0}$}
  \psfrag{CMS}[l][r][0.5]{\hspace{-0.7cm} CMS data}
  \psfrag{cos}{\raisebox{.1cm}{\scalebox{0.9}{$\langle \cos 2 \varphi\rangle / \langle \cos \varphi\rangle$}}}
  \psfrag{Y}{\scalebox{0.9}{$Y$}}
  \begin{minipage}{0.49\textwidth}
      \includegraphics[width=7cm]{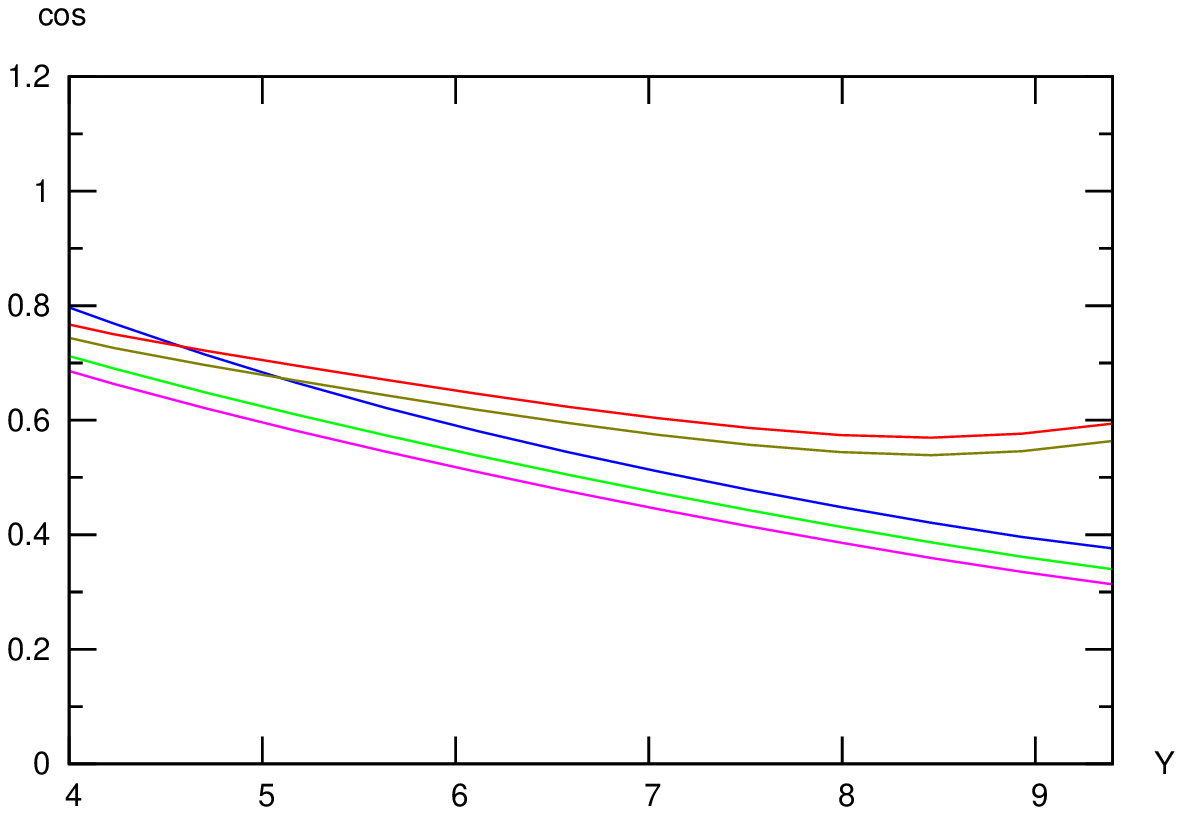}
  \end{minipage}
  \begin{minipage}{0.49\textwidth}
      \includegraphics[width=7cm]{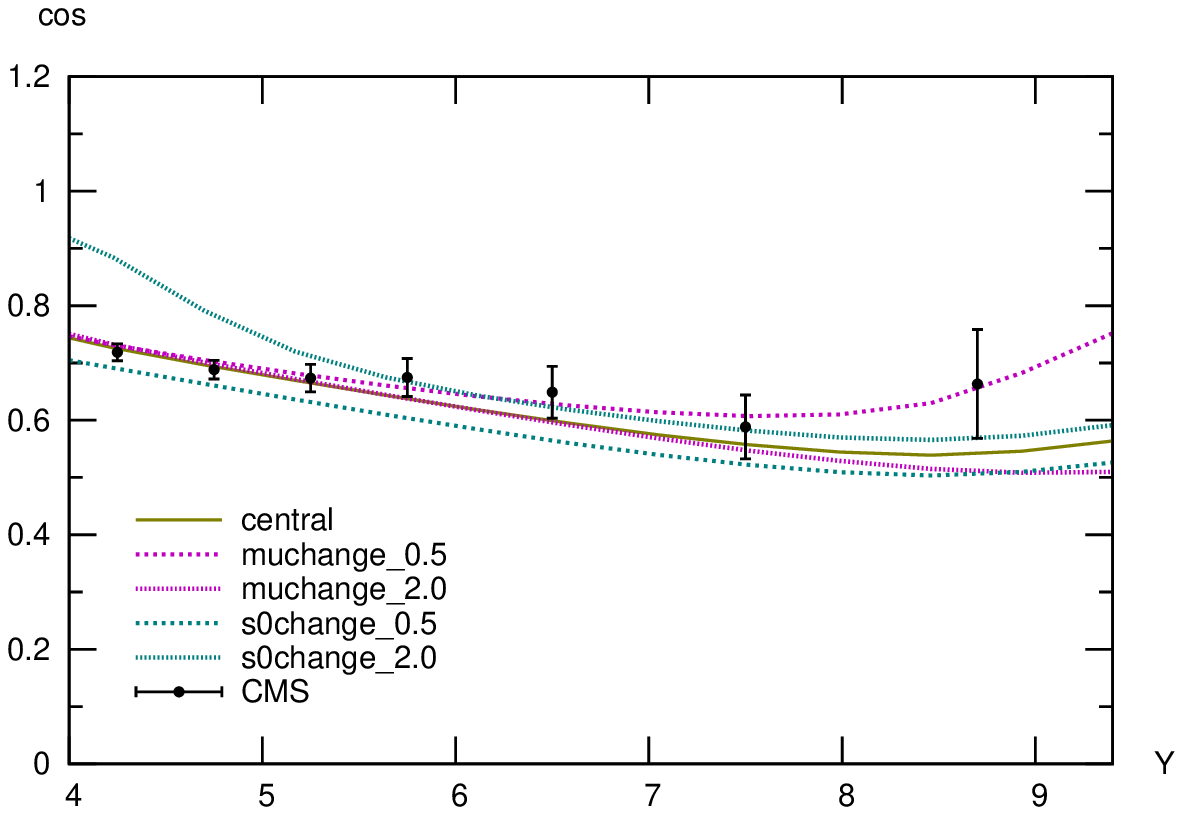}
  \end{minipage}
  \caption{Left: value of $\avgcostwo / \avgcos$ as a function of the rapidity separation $Y$, using symmetric cuts defined in (\protect\ref{sym-cuts}), for the 5 different BFKL treatments (\protect\ref{def:colors}). Right: comparison of the full NLL BFKL calculation including the scale uncertainty with CMS data.}
  \label{Fig:cos2cos_sym}
\end{figure}

The extraction of ratios of the previously mentioned observables was also performed in~\cite{CMS-PAS-FSQ-12-002}. In figure~\ref{Fig:cos2cos_sym} we show results for $\avgcostwo / \avgcos$. We see that again the effect of NLL corrections to the vertices is important, but that this observable is more stable with respect to  $\mu$ and $s_0$ than the previous ones. The agreement with data is very good over a large $Y$ range.

\section{Results: asymmetric configuration}

\begin{figure}[htbp]
  \def\sca{.6}
  \psfrag{central}[l][r][0.5]{\hspace{-0.9cm}pure NLL}
  \psfrag{muchange_0.5}[l][r][\sca]{\hspace{-1.9cm}\footnotesize $\mu_F \to \mu_F/2$}
  \psfrag{muchange_2.0}[l][r][\sca]{\hspace{-2cm} \footnotesize $\mu_F \to2 \mu_F$}
  \psfrag{s0change_0.5}[l][r][\sca]{\hspace{-2.05cm} \footnotesize $\sqrt{s_0} \to \sqrt{s_0}/2$}
  \psfrag{s0change_2.0}[l][r][\sca]{\hspace{-2.05cm} \footnotesize $\sqrt{s_0} \to 2 \sqrt{s_0}$}
  \psfrag{Dijet}[l][r][0.5]{\hspace{-0.7cm} fixed order NLO}
  \psfrag{cos}{\raisebox{.1cm}{\scalebox{0.9}{$\langle \cos \varphi\rangle$}}}
  \psfrag{Y}{\scalebox{0.9}{$Y$}}
  \begin{minipage}{0.49\textwidth}
    \hspace{-.3cm}  \includegraphics[width=7cm]{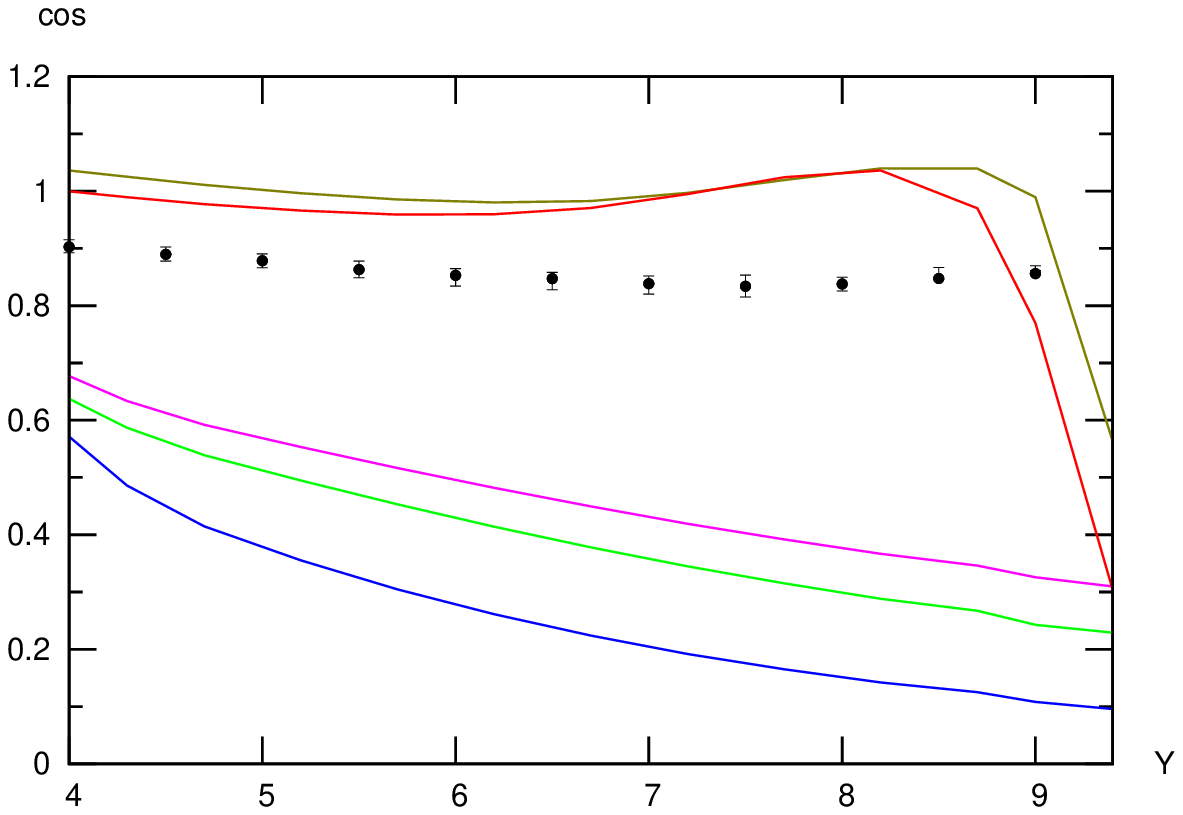}
  \end{minipage}
  \hspace{-.2cm}\begin{minipage}{0.49\textwidth}
      \includegraphics[width=7cm]{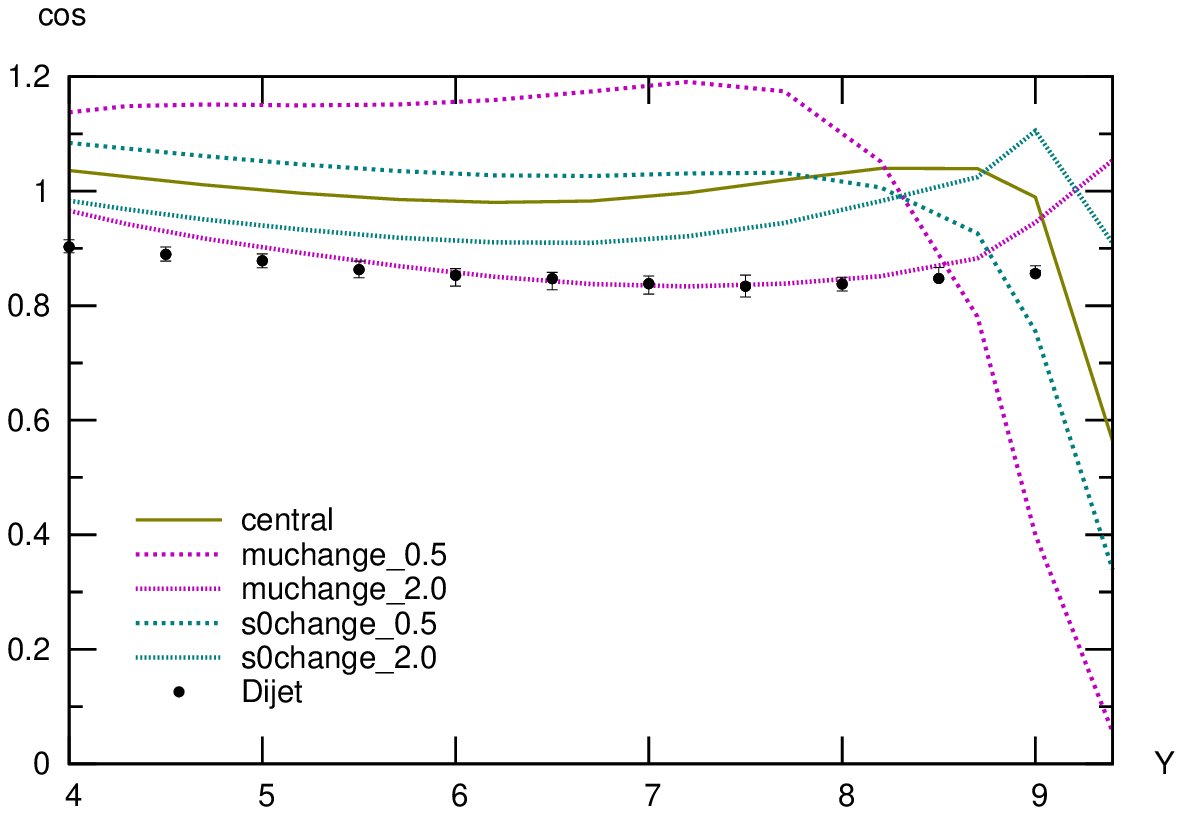}
  \end{minipage}
  \caption{Left: value of $\avgcos$ as a function of the rapidity separation $Y$, using asymmetric cuts defined in (\protect\ref{asym-cuts}), for the 5 different BFKL scenarios (\protect\ref{def:colors}). Right: comparison of the full NLL calculation including the scale uncertainty with \textsc{Dijet} predictions.}
  \label{Fig:cos_asym}
\end{figure}

To study the need for resummation to get a better description of the data, it would be interesting to study the agreement of a fixed order calculation with the data. However, fixed order calculations have instabilities when the lower cuts on the transverse momenta of the jets are identical, so we cannot compare such calculations with the results of ~\cite{CMS-PAS-FSQ-12-002}. Thus in this section we compare our BFKL calculation with the fixed order NLO code \textsc{Dijet}~\cite{Aurenche:2008dn} in an asymmetric configuration with the following cuts:
\begin{eqnarray}
  35\,{\rm GeV} < &|\veckjone|, |\veckjtwo| & < 60 \,{\rm GeV} \,, \nonumber\\
  50\,{\rm GeV} < &{\rm Max}(|\veckjone|, |\veckjtwo|)\,, \nonumber\\
 0 < &y_1, \, y_2& < 4.7\,.
 \label{asym-cuts}
\end{eqnarray}
We first consider the azimuthal correlation $\avgcos$ (figure~\ref{Fig:cos_asym}). We observe that \textsc{Dijet} predicts a much larger correlation between the jets than the three BFKL treatments using LL vertices, while a full NLL BFKL calculation produces an even larger correlation than fixed order. But when we take into account the large dependence of the BFKL calculation on the scales we find that there is an agreement between NLL BFKL and fixed order NLO.

\begin{figure}[htbp]
  \def\sca{.6}
  \psfrag{central}[l][r][0.5]{\hspace{-0.9cm}pure NLL}
  \psfrag{muchange_0.5}[l][r][\sca]{\hspace{-1.9cm}\footnotesize $\mu_F \to \mu_F/2$}
  \psfrag{muchange_2.0}[l][r][\sca]{\hspace{-2cm} \footnotesize $\mu_F \to2 \mu_F$}
  \psfrag{s0change_0.5}[l][r][\sca]{\hspace{-2.05cm} \footnotesize $\sqrt{s_0} \to \sqrt{s_0}/2$}
  \psfrag{s0change_2.0}[l][r][\sca]{\hspace{-2.05cm} \footnotesize $\sqrt{s_0} \to 2 \sqrt{s_0}$}
  \psfrag{Dijet}[l][r][0.5]{\hspace{-0.7cm} fixed order NLO}
  \psfrag{cos}{\raisebox{.1cm}{\scalebox{0.9}{$\langle \cos 2 \varphi\rangle$}}}
  \psfrag{Y}{\scalebox{0.9}{$Y$}}
  \begin{minipage}{0.49\textwidth}
    \hspace{-.3cm}  \includegraphics[width=7cm]{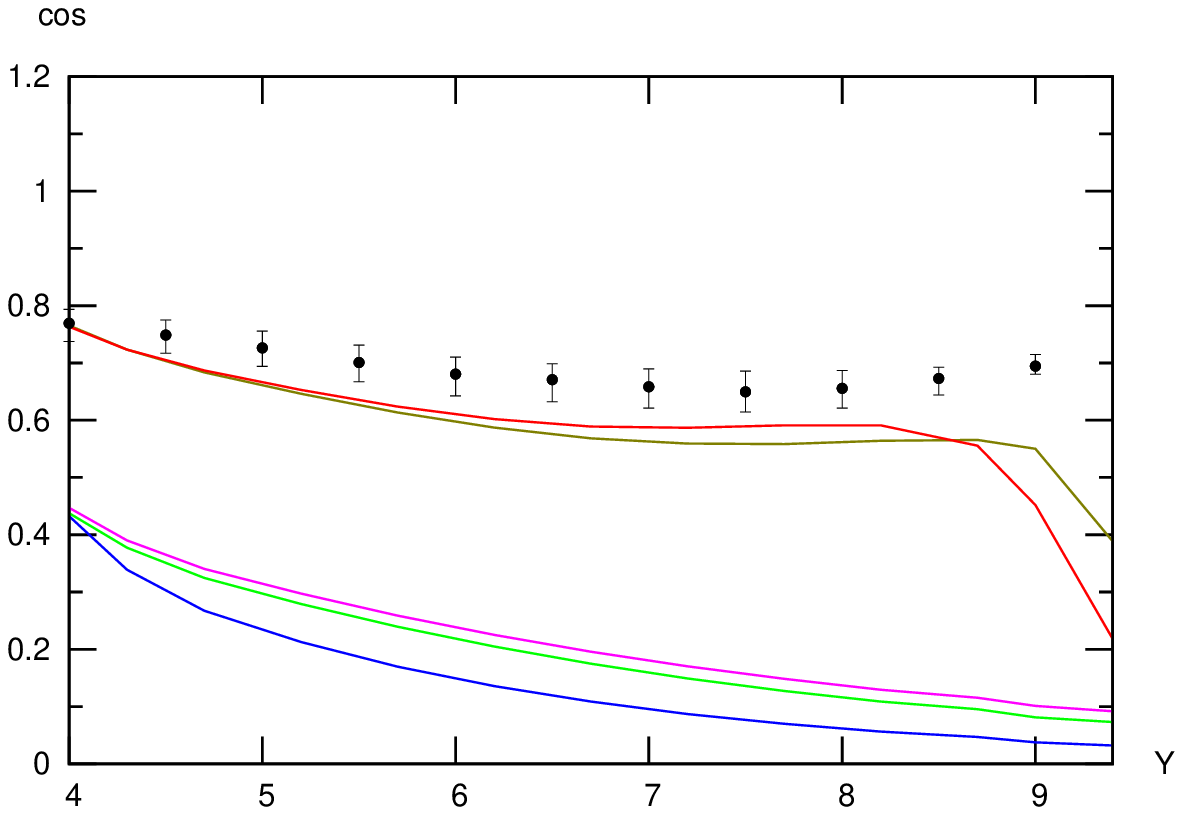}
  \end{minipage}
  \begin{minipage}{0.49\textwidth}
    \hspace{-.3cm}  \includegraphics[width=7cm]{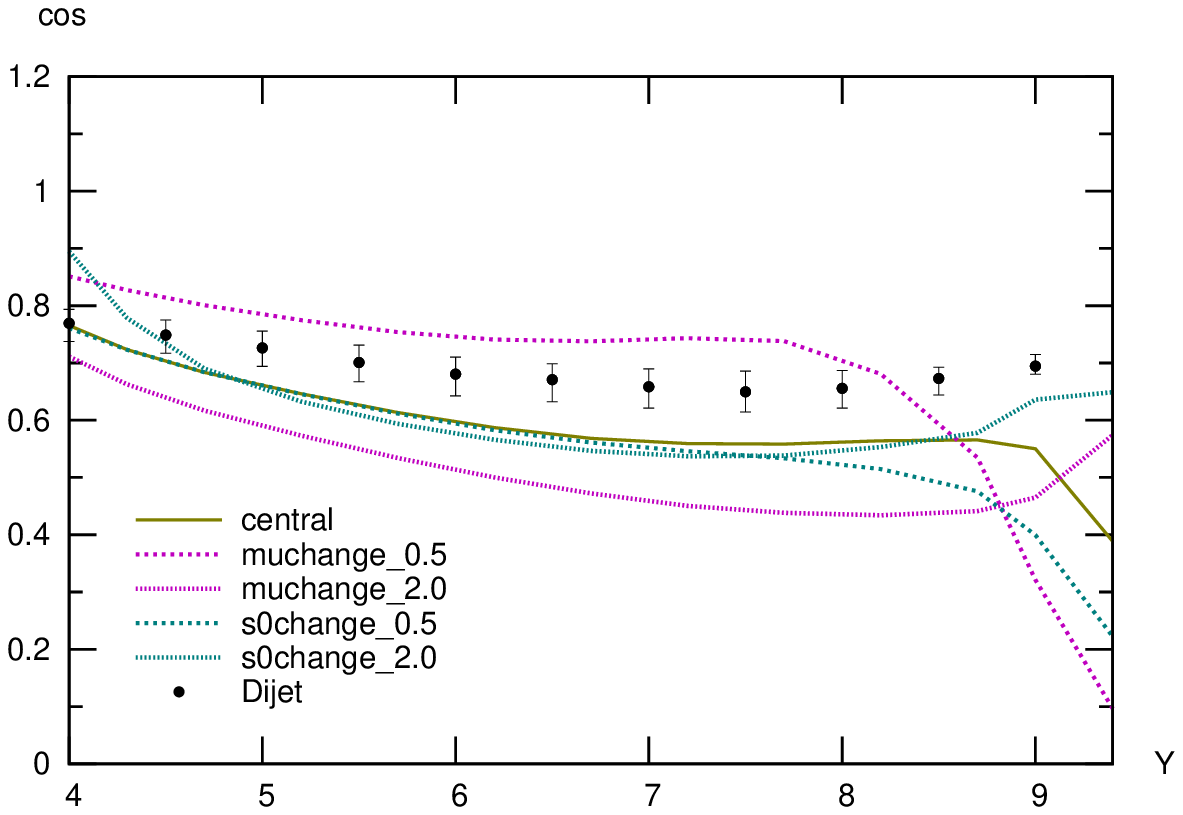}
  \end{minipage}
  \caption{Left: value of $\avgcostwo$ as a function of the rapidity separation $Y$, using asymmetric cuts defined in (\protect\ref{asym-cuts}), for the 5 different BFKL scenarios (\protect\ref{def:colors}). Right: comparison of the full NLL calculation including the scale uncertainty with \textsc{Dijet} predictions.}
   \label{Fig:cos2_asym}
\end{figure}

A similar conclusion can be drawn when considering $\avgcostwo$, as shown in figure~\ref{Fig:cos2_asym}. Again the scale uncertainty does not allow to distinguish between NLL BFKL and fixed order NLO.

\begin{figure}[htbp]
  \def\sca{.6}
  \psfrag{central}[l][r][0.5]{\hspace{-0.9cm}pure NLL}
  \psfrag{muchange_0.5}[l][r][\sca]{\hspace{-1.9cm}\footnotesize $\mu_F \to \mu_F/2$}
  \psfrag{muchange_2.0}[l][r][\sca]{\hspace{-2cm} \footnotesize $\mu_F \to2 \mu_F$}
  \psfrag{s0change_0.5}[l][r][\sca]{\hspace{-2.05cm} \footnotesize $\sqrt{s_0} \to \sqrt{s_0}/2$}
  \psfrag{s0change_2.0}[l][r][\sca]{\hspace{-2.05cm} \footnotesize $\sqrt{s_0} \to 2 \sqrt{s_0}$}
  \psfrag{Dijet}[l][r][0.5]{\hspace{-0.7cm} fixed order NLO}
  \psfrag{cos}{\raisebox{.1cm}{\scalebox{0.9}{$\langle \cos 2 \varphi\rangle / \langle \cos \varphi\rangle$}}}
  \psfrag{Y}{\scalebox{0.9}{$Y$}}
  \begin{minipage}{0.49\textwidth}
  \hspace{-.3cm}  \includegraphics[width=7cm]{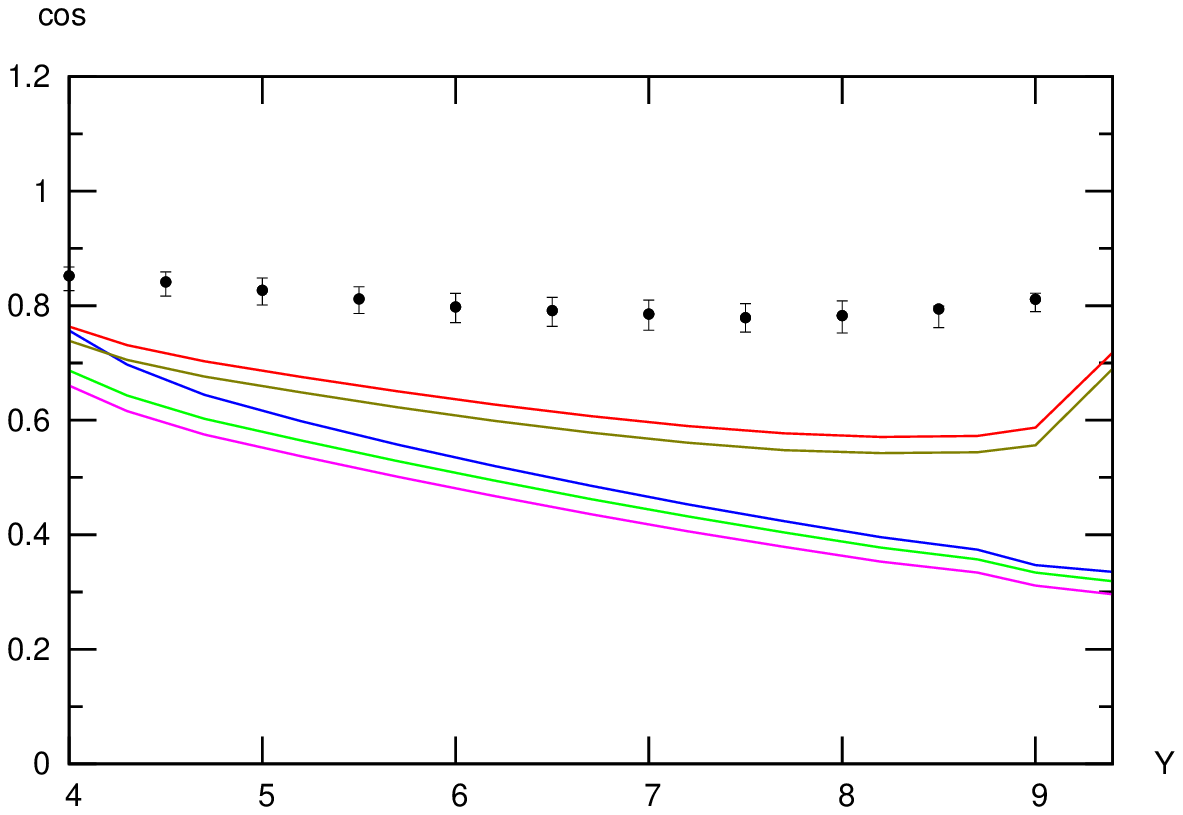}
  \end{minipage}
  \begin{minipage}{0.49\textwidth}
  \hspace{-.3cm}    \includegraphics[width=7cm]{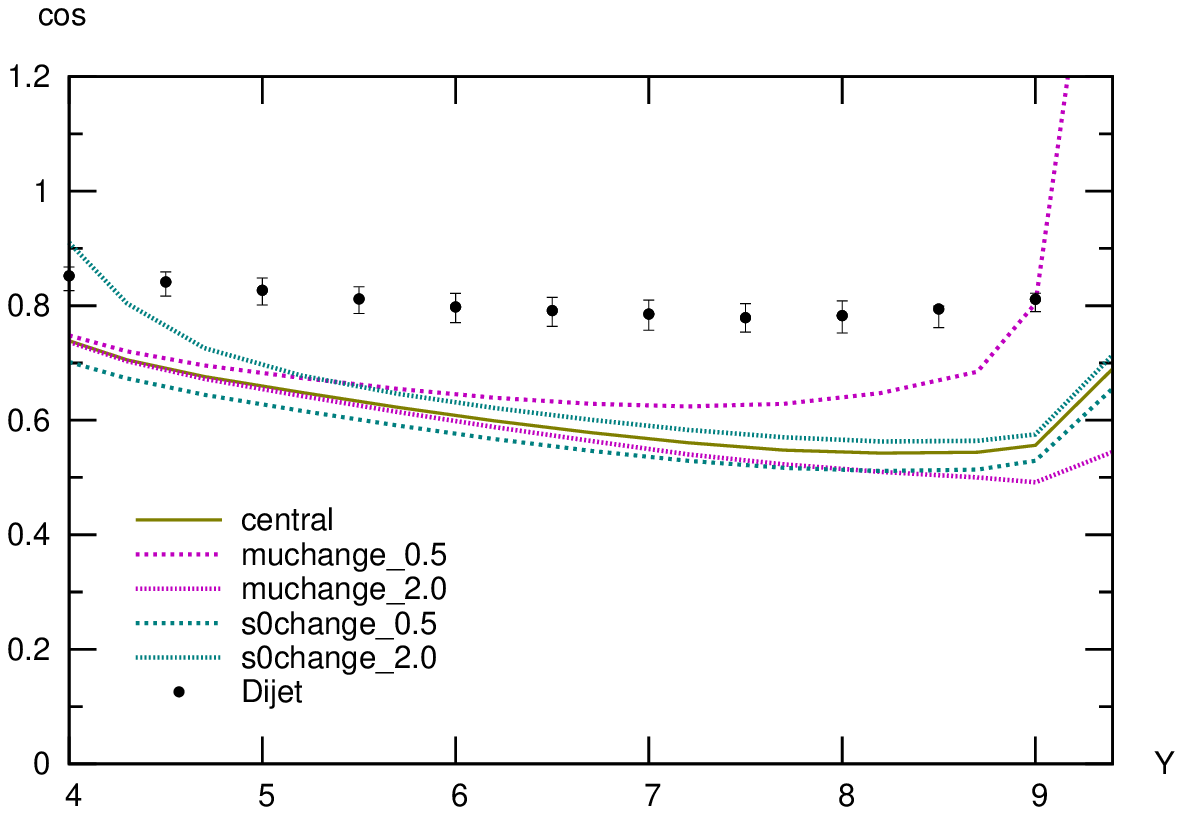}
  \end{minipage}
   \caption{Left: value of $\avgcostwo / \avgcos$ as a function of the rapidity separation $Y$, using asymmetric cuts defined in (\protect\ref{asym-cuts}), for the 5 different BFKL scenarios (\protect\ref{def:colors}). Right: comparison of the full NLL calculation including the scale uncertainty with \textsc{Dijet} predictions.}
\label{Fig:cos2cos_asym}
\end{figure}

In figure~\ref{Fig:cos2cos_asym} we show results for the observable $\avgcostwo / \avgcos$. Here the fixed order NLO calculation is significantly above all the BFKL calculations. As in the symmetric case this observable is quite stable with respect to the scales so the difference between NLL BFKL and fixed order NLO does not vanish when we take into account the scale uncertainty.

\section{Conclusions}

For the first time, we have been able to compare the predictions of our full NLL BFKL calculation of Mueller-Navelet jets with data taken at the LHC thanks to data presented by the CMS collaboration. This comparison shows that for the observables $\avgcosn$ a pure LL BFKL treatment or a mixed treatment where the NLL Green's function is used together with LL vertices cannot describe the data. On the other hand, the results of our complete NLL calculation do not agree very well with the data when the scales involved are fixed at their 'natural' value $\sqrt{|\veckjone|\cdot |\veckjtwo|}$, but as the dependence on these scales is still quite large no firm conclusion can be drawn at the moment. The investigation of these issues is essential and left for future work.
On the contrary we saw that ratios of these observables are more stable with respect to changes of the scales and describe the data quite well.

To find an evidence for the need of BFKL-type resummation a comparison with a fixed order treatment would be needed to see if the BFKL calculation provides a better description of the data. For the moment we cannot do such comparison as the configuration chosen by the CMS collaboration would lead to unstable results in a fixed order calculation. However, we compared our results with the fixed order NLO code \textsc{Dijet} in an asymmetric configuration and found that for the observables $\avgcosn$ no significant difference is observed when taking into account the uncertainties associated with the choice of the scales. In contrast we see that for $\avgcostwo / \avgcos$ the two calculations lead to noticeably different results. This, added to the fact that this observable is quite stable with respect to the scales, confirms that it seems to be well-suited to study resummation effects at high energy and that an experimental analysis with slightly different lower cuts on the transverse momenta of the jets 
may be of interest.

\acknowledgments

We thank Michel Fontannaz, Cyrille Marquet and Christophe Royon for
providing their codes and for stimulating discussions.
We warmly thank Grzegorz Brona, David d'Enterria, Hannes Jung, Victor Kim and
Maciej Misiura for many discussions and fruitful suggestions on the experimental aspects
of this study.

This work is supported by
the Polish Grant NCN No.~DEC-2011/01/B/ST2/03915 and  the Joint Research Activity Study of Strongly Interacting Matter (HadronPhysics3, Grant Agreement n.283286) under the 7th Framework Programm of the European Community.

\providecommand{\href}[2]{#2}\begingroup\raggedright\endgroup

\end{document}